# Observations with small and medium-sized telescopes at the Terskol Observatory

**V. Tarady, O. Sergeev, M. Karpov, B. Zhilyaev, V. Godunova**

International Center for Astronomical, Medical, and Ecological Research
27 Zabolotnoho Str., 03680 Kyiv, Ukraine
*e-mail: godunova@mao.kiev.ua*

**Abstract** Astronomical facilities at the high-altitude observatory Terskol in the Northern Caucasus include optical telescopes with diameters up to 2 m, their instrumentation (high-resolution spectrometers, high-speed photometers, CCDs, etc.), as well as provisions for data distribution via satellite and computer networks. The decades of successful research at Terskol have yielded new data and findings in the following areas of astronomy: discovery and monitoring of NEOs, precise astrometry and photometry of solar system bodies, high-resolution spectroscopy of interstellar clouds, search for optical afterglow of gamma ray bursts, etc. Facilities of the Terskol Observatory are heavily used for the operation of the Synchronous Network of distant Telescopes, which includes optical telescopes at Terskol and at observatories in Bulgaria, Greece, and Ukraine; the remarkable results were obtained especially from synchronous observations of galaxies and flare stars.

## 1 Introduction

Construction of an observatory on Terskol Peak in the Northern Caucasus ($43^{\circ}16'29''$ N, $42^{\circ}30'03''$ E) began in the 1970s. During the next two decades, this site was developed as a high-altitude observation station of the Main Astronomical Observatory of the Ukrainian Academy of Sciences. Since 1993, the Terskol Observatory has been operated by the International Center for Astronomical, Medical, and Ecological Research (ICAMER), a joint institution of the Ukrainian and Russian academies of sciences.

Terskol Peak is a small plateau at an altitude of 3100 m above sea level on the southern slope of Mt Elbrus (Fig. 1). Due to high atmospheric transparency, this site is well suited for astronomical research in the UV wavelength range. Moreover, the low water vapor content in the atmosphere over Terskol provides a unique environment for IR observations. There are about 160 clear nights per year with a seeing of about 1 arcsec.

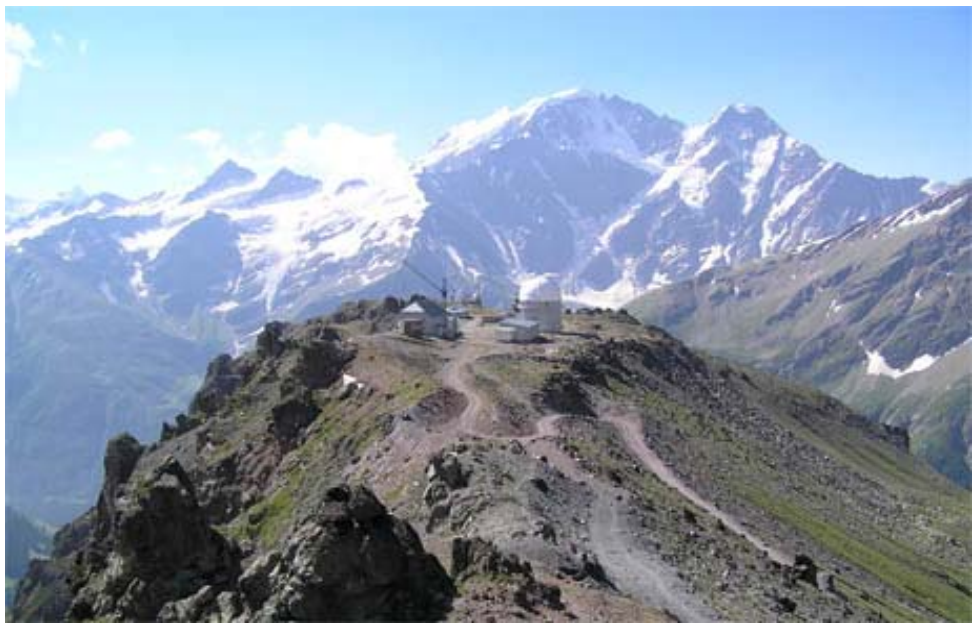

**Fig. 1** A panoramic view of the Terskol Observatory from the North

**2 Observational facilities: status and developments**

2.1 2-m Ritchey-Chretien-Coude telescope

The main instrument of the Terskol Observatory is a 2-m Ritchey-Chretien-Coude telescope. It is the fifth telescope of this size manufactured by Carl Zeiss Jena GmbH. It was put into operation in 1995 (Fig. 2). The focal length of the main hyperbolic mirror is 5.6 m. The equivalent focal lengths of the Ritchey-Chretien and Coude systems are 16 m and 72 m, respectively. The corresponding fields of view are 108 arcmin and 5 arcmin.

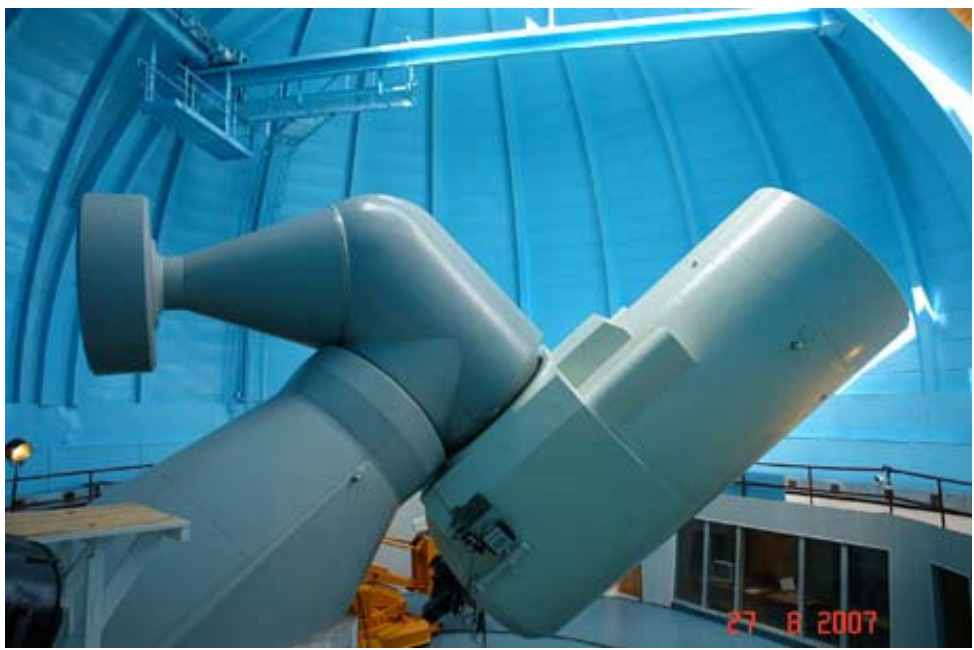

**Fig. 2** The 2-m Ritchey-Chretien-Coude telescope

The photometric complex of the 2-m telescope includes a two-channel high-speed photometer with cooled photo-multipliers, UBVRI filters and a CCD guiding system. This complex has a precise timing and synchronization system based on the GPS smart antenna Acutime-2000. The accuracy of the timing and synchronization is better than l ms.

The additional instrumentation of the telescope consists of high-efficiency CCD cameras, spectrographs and some specific photometers (see section 3).

In 2008, the original control system of the telescope, VILATI, was replaced by a new one based on industrial automation tools. This system was installed by ProjectSoft HK a.s. (Czech Republic).

The 2-m telescope remains the main instrument of the Terskol Observatory and continues to make a valuable contribution to studying the Universe.

2.2 Solar telescopes

During 1986–1992, absolute measurements of the solar disk-centre intensity were performed at Terskol. The observations were carried out using the instrumentation adapted for quasi-simultaneous registration of the solar and standard ribbon tungsten lamp spectra (Burlov-Vasiljev et al. 1995, 1998). The following facilities installed at Terskol were involved:

- **grating spectrophotometer** with collimator ($D = 180$ mm, $f = 2$ m), camera mirror ($D = 230$ mm, $f = 2$ m), and grating (140 × 150 mm, 600 or 1200 grooves/mm) mounted in the vertical plane

- **absolute calibration system**, which uses a ribbon tungsten lamp placed at the focus of the collimating mirror ($D = 230$ mm, $f = 3$ m)

- **atmosphere halo photometer** used for monitoring of the optical properties of the Earth's atmosphere

- **auxiliary equipment** to control the apparatus includes optical and electronical stuff for lamp feeding and control, for the investigations of the polarization properties of the coelostat group and spectrograph, for the measurements of spectral reflectivity of the lamp's collimator, etc.
- **solar horizontal telescope** with coelostat group (main mirror $D$ = 230 mm, $f$ = 3 m).

As a result, the absolute spectral energy distribution was measured for the spectral interval 305 nm - 1070 nm (Fig. 3). Moreover, it was shown that the calibration for the ground-based measurements can be fulfilled much more precisely and reliably than for the space-based ones (Burlov-Vasiljev et al. 1997).

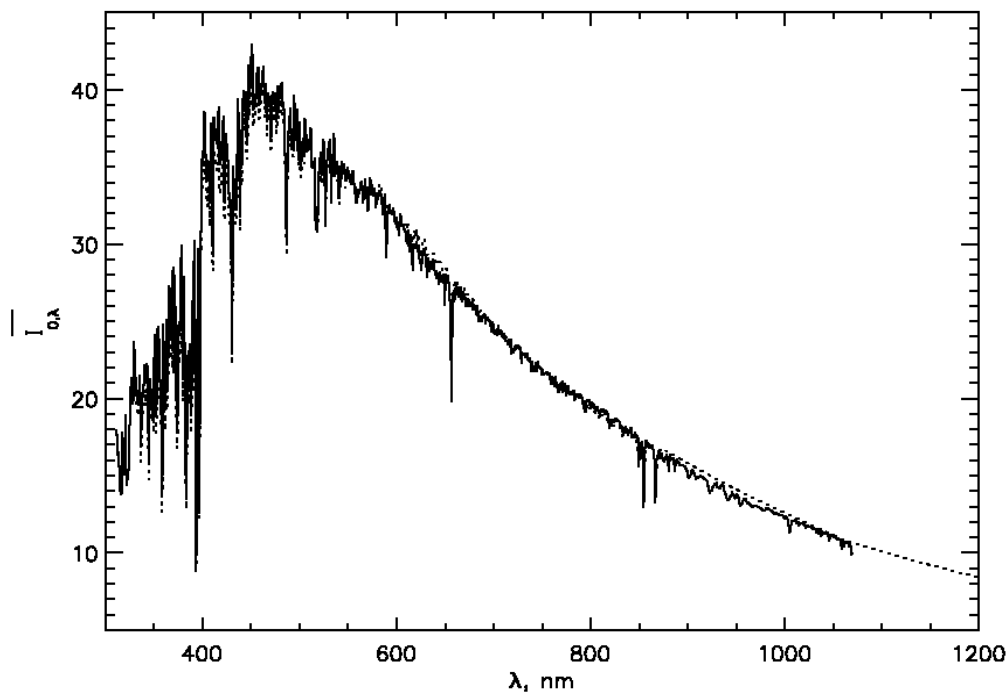

**Fig. 3** The measured spectral (continuum + lines) intensity of the solar disk-center radiation in $10^{12}$ W·m$^{-3}$·ster$^{-1}$ (Burlov-Vasiljev et al. 1997)

In 1992, the large horizontal solar telescope, ACU-26, built by the Optical-Mechanical Corporation LOMO (Russia), was put into operation (Fig. 4). The diameter of the main spherical mirror is 650 mm; its focal length is 17.75 m. The telescope is equipped with a 5-camera Ebert-Fastie spectrograph (Fig. 5).

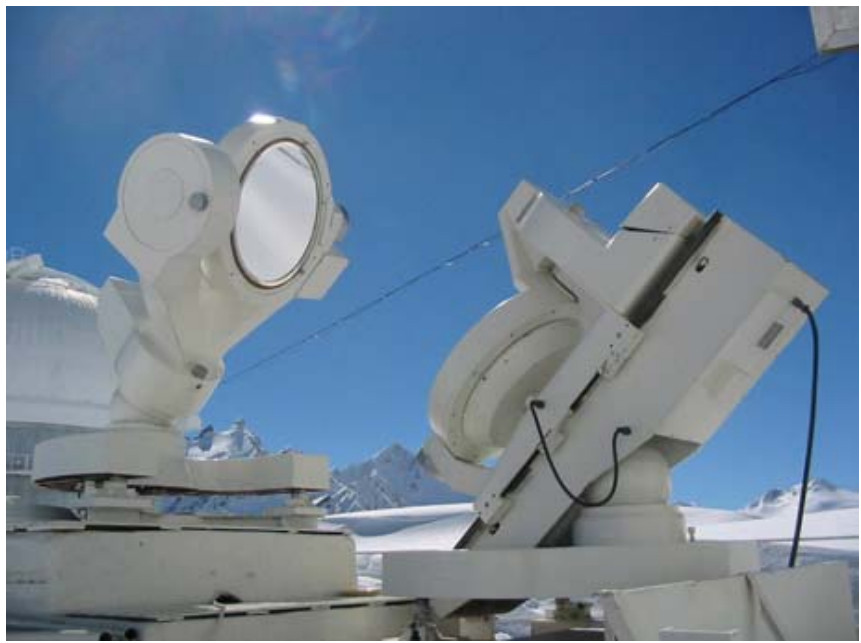

**Fig. 4** The large horizontal solar telescope ACU-26

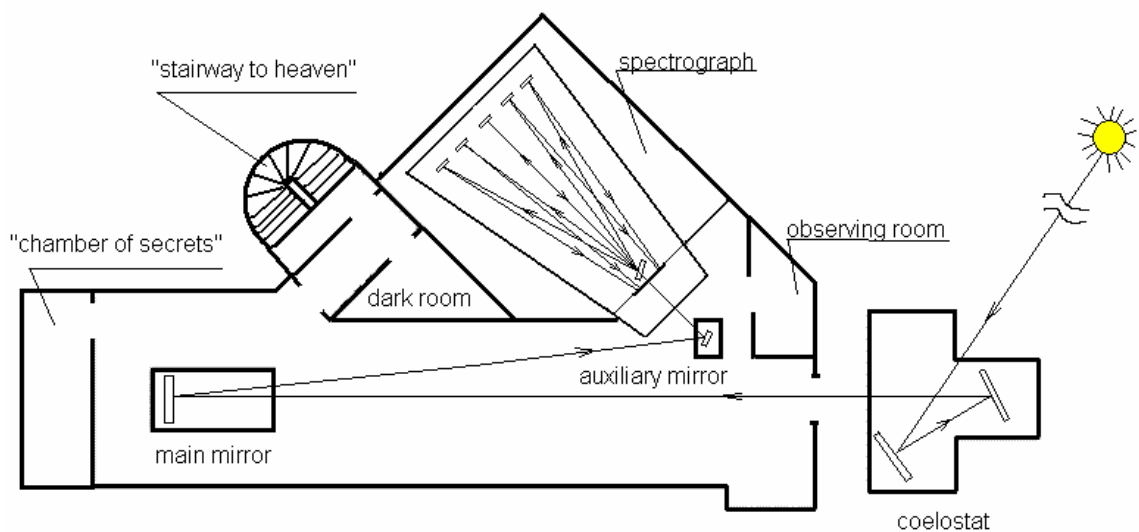

**Fig. 5** The scheme of the large horizontal solar telescope ACU-26

The diameter of the spectrograph's collimator and mirrors is 300 mm; the focal length of the collimator and mirrors is 8 m. The grating size is 250 x 200 mm with 600 l mm$^{-1}$. The dispersions in the fourth order are 21.9 mm/nm at 395.0 nm and 33.0 mm/nm at 650.0 nm. The typical instrumental profile of the spectrograph has a FWHM of 0.0018 nm.

The telescope is useable to perform observations of selected regions or active areas on the Sun simultaneously in five wavelength ranges.

2.3 The Zeiss-600 telescope

The 60-cm Cassegrain telescope (Zeiss-600) has been in operation since the mid-1970s (Fig. 6). Its focal length is 7.5 m. Depending on the CCD camera attached the field of view is 9 x 8 or 4 x 3 arcmin. This telescope is currently used for astrometric observations with timing better than 1 ms, as well as for photometric observations with B, V, G, I or R filters. Observational programmes are aimed at obtaining precise data sets to study variable stars and asteroids.

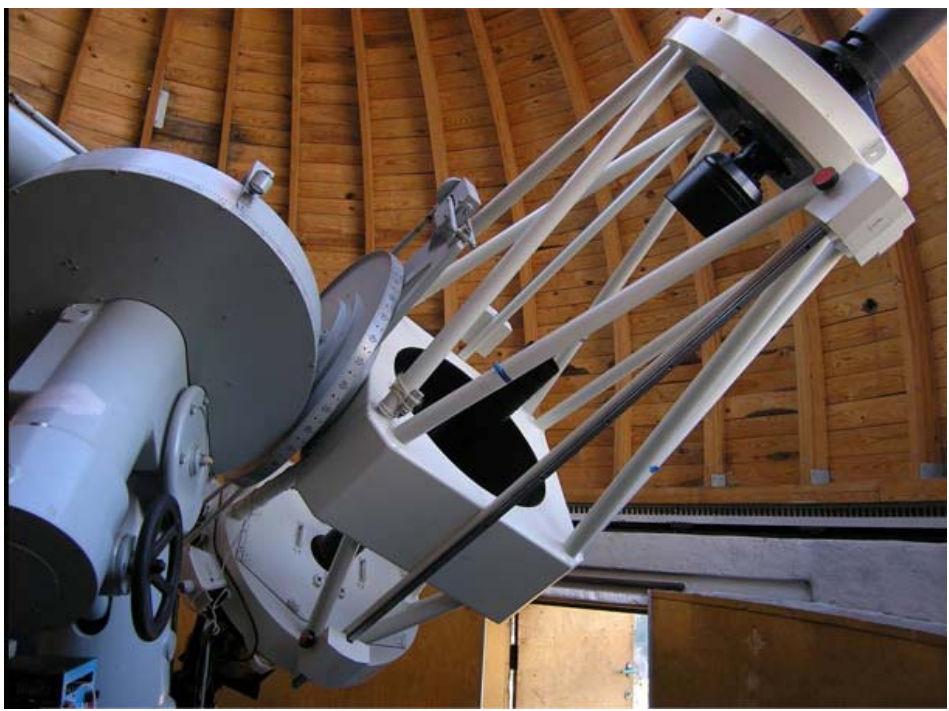

**Fig. 6** The Zeiss-600 telescope

2.4 Other telescopes and tools

Included among the telescopes installed at the Terskol Observatory are a Celestron 11", a Meade 14", and an 80-cm telescope. But we still need to do some more work to start comprehensive scientific programmes with these telescopes.

During the last decade, the infrastructure to support observational activities at Terskol has been updated considerably. For instance, a DIRECWAY $^{TM}$ system is now used for Internet connectivity via satellite through a satellite antenna. Furthermore, to provide a proper maintenance of the astronomical instruments, a liquid nitrogen generator and turbo-molecular pumps are available on Terskol Peak.

## 3 Further steps in designing and using additional instruments at Terskol

During the past few decades, the situation in ground-based and space astronomy has changed dramatically, to a certain degree thanks to new observational facilities. However, not only new telescopes and new satellites or planet missions give the possibility to answer some of the fundamental questions. Many advances in astronomy come from the development and use of specific instruments and techniques. Thus, ground-based telescopes (including small- and medium-aperture telescopes, equipped with CCD cameras and additional instruments) still provide good enough opportunities for long-term astrometric, photometric, polarimetric, and other observations.

3.1 Exploration of Solar System bodies

*3.1.1 The focal reducer for high-quality imaging*

Observations of Solar System bodies started at Terskol in 1996 with the aid of the two-channel focal reducer attached to the Cassegrain focus of the 2-m telescope. This instrument was built by the Max-Planck Institute for Solar System Research (formerly Max-Planck Institute for Aeronomy - MPAe) and brought to the Terskol Observatory according to the agreement on cooperation between MPAe and ICAMER.

The focal reducer is a lens system consisting of collimator, parallel beam and camera lens with high light gathering power. It shrinks the telescope image to a size that matches the seeing disk and the required field of view with the detector (CCD) resolution. This property makes the reducer suitable for taking atmospheric turbulence into account (Jockers 1997).

Atmospheric turbulence is one of the factors limiting the magnification of optical ground-based telescopes. With the 2-m telescope at Terskol it is impossible to get a resolution much better than 1/20000 of a degree, i. e. 0.18 arcsec. However, the focal length of the telescope can be adjusted to the task of the observations.

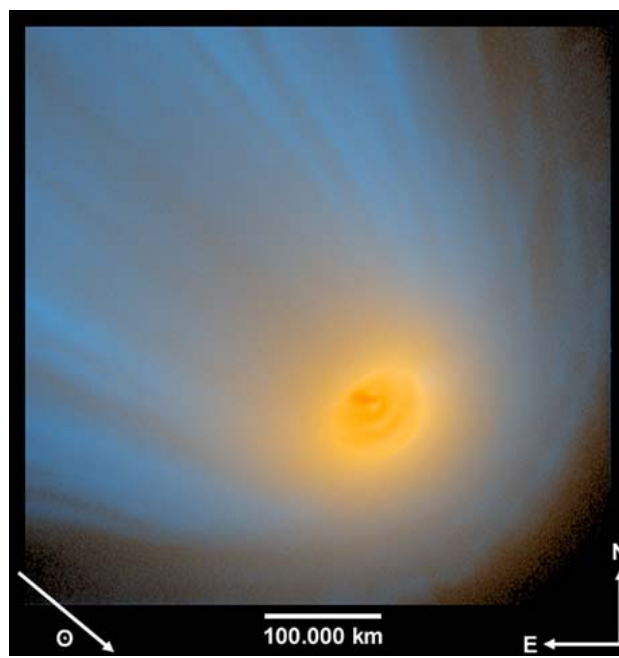

**Fig. 7** Two-color image of comet C/1995 O1 (Hale-Bopp) obtained at Terskol with the Two-Channel Focal Reducer on April 13, 1997. The orange color represents the dust grains of the cometary atmosphere. Note the dust spirals observed in this huge comet. The blue color represents the distribution of cometary ions (here the ion $OH^+$) (Jockers 2003)

In its Cassegrain focus, the 2-m telescope has a focal length optimized to the best conditions of atmospheric turbulence. The two-channel focal reducer reduces this focal length by a factor of 2.86 to a resolution of about 1 arcsec. This allows one to obtain sharp images even under non-optimum conditions of atmospheric turbulence. At the same time the light-gathering power increases by the square of the same factor ($2.86^2 = 8.2$), i. e. less time is needed to get a well-exposed image (Jockers 2003).

As an additional feature, the focal reducer has two channels to conduct simultaneous observations in two wavelength bands between 350 nm and 1000 nm. Moreover, it is equipped with Fabry-Perot etalons for narrow-band observations.

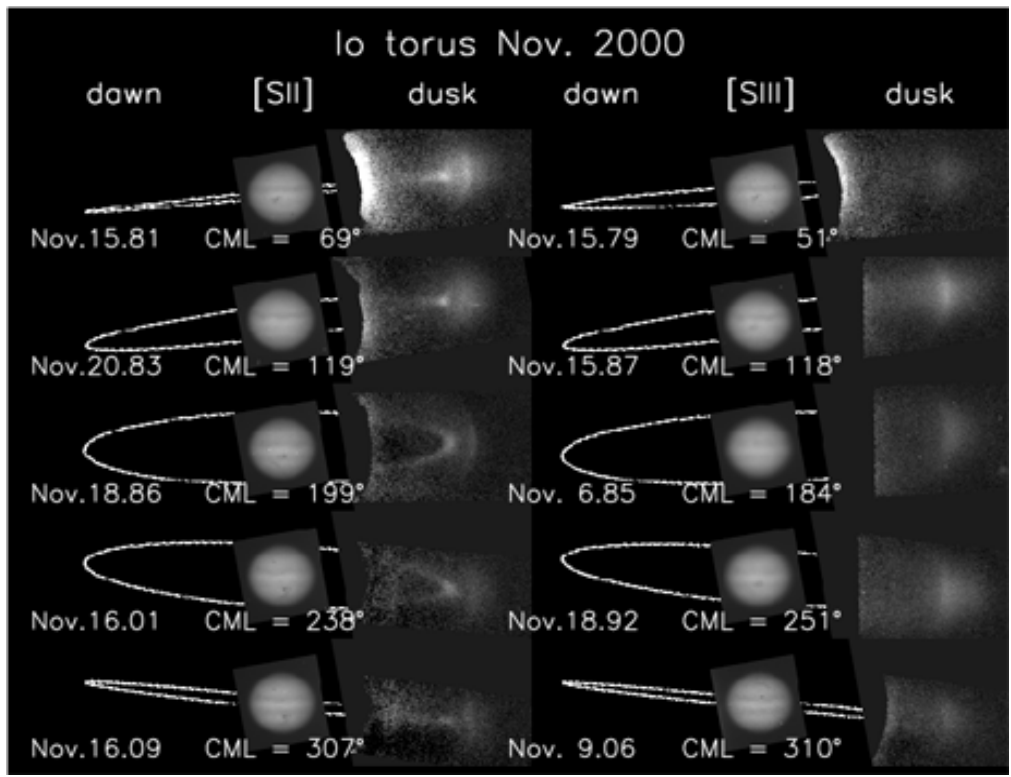

**Fig. 8** The Io torus, imaged in the forbidden lines of $S^+$ ([SII]) and $S^{++}$ ([SIII]) (Jockers 2003)

Using this electronic imaging device, in 1996-2002 the MPAe researchers (Prof. Klaus Jockers and his group), in collaboration with Russian, Ukrainian and Bulgarian astronomers, successfully studied gas and dust in comets, conducted polarimetry of cometary dust and asteroids, and astrometry and photometry of the inner Jovian satellites (Jockers et al. 1998; Kulyk & Jockers 2004). Figure 7 shows a two-color image of comet C/1995 O1 (Hale-Bopp) obtained with the two-channel focal reducer at Terskol. Moreover, the morphology and brightness of the plasma torus of the Jovian satellite Io were investigated in order to derive its physical properties (Fig. 8).

In recent years, a new focal reducer based on the original Meinel camera, was developed and constructed at the Terskol Observatory.

*3.1.2 Multi-Mode Cassegrain Spectrograph and other instruments to support research activities*

The Multi-Mode Cassegrain Spectrograph (MMCS) was designed in 2003 for observations of faint objects at the Cassegrain focus (F/8) of the 2-m telescope. It has two modes of functioning: (i) grating / echelle spectrometer mode (300 nm -1200 nm, R =1000…15000, the limiting magnitude is ~15.5 mag); (ii) CCD photometer mode.

The main components of MMCS are as follows:
- collimator is a parabolic mirror with diameter of 75 mm and focal length of 600 mm;
- special (with a central hole) echelle grating with 75 gr mm$^{-1}$ and 63. 5° blaze angle;
- diffraction grating with 600 gr mm$^{-1}$ and 8° blaze angle;
- diffraction grating with 300 gr mm$^{-1}$ and 4° blaze angle;
- diffraction grating with 300 gr mm$^{-1}$ and 6° blaze angle;
- diffraction grating with 200 gr mm$^{-1}$ and 28° blaze angle;

- a 45° crown prism;
- Schmidt-Cassegrain camera with F=150 mm;
- lens camera with F=180 mm;
- detectors: Wright Instruments CCD, 1242x1152 pixels of 22.5 μm; Wright Instruments CCD (back-illuminated), 1252x1152 pixels of 22.5 μm.

Some of the observational capabilities of MMCS are presented in Table 1.

**Table 1** Some observational capabilities of The Multi-Mode Cassegrain Spectrograph

| Modes | echelle (64°+4°) | quasi-echelle (28°+prism) | classic 8°-grating | classic 4°-grating |
|---|---|---|---|---|
| Maximum resolution | 13500 | 3200 | 1200 | 600 |
| Limiting magnitude, S/N~10, $1^h$ exposure | ~12.5 | ~14.5 | ~15 | ~16 |

MMCS and some specific automatic photometers, which were developed and constructed at the Terskol Observatory, are in productive scientific use on the telescopes. These devices contribute significantly to achieving important results in the following fields of research:
- detection and monitoring of potentially hazardous objects (Earth-approaching asteroids, comets) and space debris,
- precise astrometry and photometry of comets and asteroids,
- photometric observations of trans-neptunian objects (Rousselot et al. 2005),
- high-resolution mapping of planetary surfaces by the short exposure method.

Moreover, the abovementioned instruments are heavily used within other observational programmes, which have been run at Terskol: photometry of variable stars and cataclysmic variables (also within the framework of the Whole Earth Telescope), search for optical afterglow of gamma ray bursts, etc.

The program on NEOs observations is carried out in collaboration with the Institute of Astronomy (Russia) and includes an astrometric position determination and taxonomic investigation of the objects. The technique applied allows us to detect potentially hazardous bodies of decameter size at a distance of some million kilometers. Using photometric and spectroscopic observations, a larger number of asteroids were classified with regard to size and taxonomic type.

As for observations of satellites in or near geosynchronous orbit, several important results were also obtained. For instance, a critical approach of the geosynchronous satellite Arabsat 1C to a Russian satellite was revealed in February 1997.

3.2 Up-to-date science with a high-resolution spectrometer

Important advances in observational techniques, data acquisition and processing have been made at the Terskol Observatory when in the late 1990s an echelle spectrometer was developed and put into operation. This work was performed in collaboration with Prof. Jacek Krelowski and his colleagues from the Toruń Center for Astronomy of the Nicolaus Copernicus University (Poland).

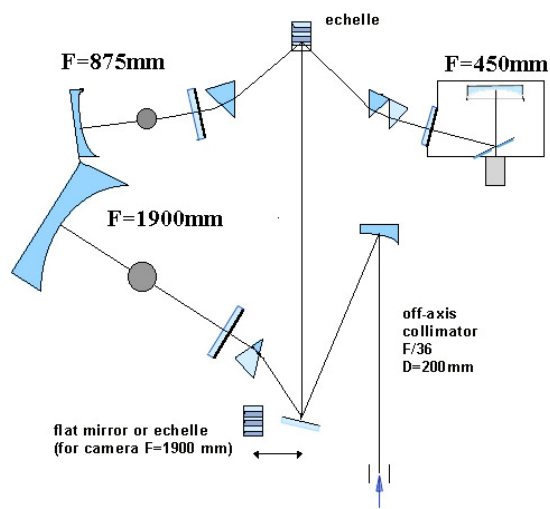

**Fig. 9** The schematic layout of the 3-camera cross-dispersed echelle spectrometer

The spectrometer called MAESTRO (MAtrix Echelle SpecTROmeter) was completed and installed in an isolated and temperature-stable coude-room in the tower of the 2-m telescope. It offers a range of resolutions: R=$\lambda/\delta\lambda$ = 45,000; 120,000; 210,000 and 500,000. These resolutions can be achieved by means of three Schmidt cameras with focal lengths of 450 mm, 875 mm and 1960 mm, respectively (Fig. 9). The limiting magnitude of the spectrometer is about $11^m$ (S/N ~ 10, $1^h$ exposure).

The main components of the cross-dispersed echelle spectrometer MAESTRO are as follows:
- collimator is an off-axis parabolic mirror with diameter of 200 mm and focal length of 7100 mm;
- mosaic R2: built from two 200x250 mm echelle gratings with 37.5 gr mm$^{-1}$ and 63.5° blaze angle;
- mosaic R6: built from three 220x320 mm echelle gratings with 37.5 gr mm$^{-1}$ and 80.5° blaze angle;
- cross-disperser is a 45° crown prism;
- cameras: 1 - Schmidt (folded), $f$ = 450 mm; 2 - Schmidt with outer focus $f$= 875 mm; 3 - Schmidt with outer focus $f$ = 1900 mm;
- detectors: Wright Instruments CCD, 1242x1152 pixels of 22.5 μm; Wright Instruments CCD (back-illuminated), 1252x1152 pixels of 22.5 μm.

The spectrometer attached to the 2-m telescope opens up entirely new fields of research. For instance, absorption spectra of dark interstellar clouds reveal the complexity of physicochemical processes inside them. Some observed interstellar absorption features, especially diffuse interstellar bands (DIBs), still remain unidentified. An analysis of their high-resolution profiles seems to be the most prospective way to identify their carriers. Since the Terskol Observatory is located at a remote and high-altitude mountain site there are a very dark sky background and the seeing better than anywhere in Europe. The very high resolution (up to R=500000) achieved allows analysis of profile shapes of interstellar spectral features. The first spectra acquired at Terskol with the aid of the MAESTRO spectrometer clearly showed that the substructure of DIBs can be easily traced in these data. The profile details matched perfectly with those observed at other observatories (Fig. 10).

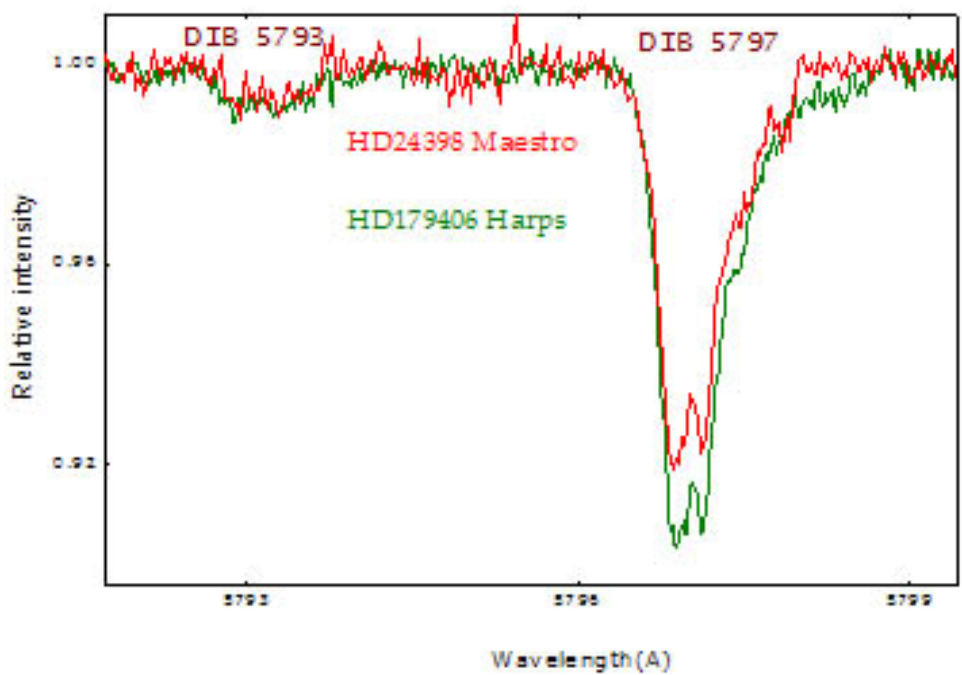

**Fig. 10** Profiles of the 5797 DIB acquired for the target HD179406 with the aid of the HARPS spectrograph at ESO and for HD24398 using the MAESTRO at Terskol (R=500000). The shapes of the two profiles are clearly identical; all substructure details are seen in both.

It should also be mentioned that the Terskol echelle spectra allow a very precise (up to 0.0003 nm) wavelength determination of any of the detectable features due to the application of a global dispersion curve.

With the aid of the echelle spectrometer installed at the coude focus of the 2-m telescope, various studies of interstellar spectral features have been conducted at Terskol. The spectrometer is capable of recording the infrared spectra of homonuclear molecules such as C2 (Phillips 2-0 band) or C3; the first images of the Phillips 2-0 band with the resolution of R=120000 were those from MAESTRO (Krelowski et al. 2003). Another important observational result was evidence for the existence of the neutral (independent of wavelength) interstellar absorption.

In 2007, the UV branch of the MAESTRO spectrograph was developed. At present, the spectral region to be covered is from 300 nm to 1000 nm. This allows us to observe OH and NH molecular bands, as well as TiII lines in order to extend our constraints to the chemistry of oxygen and nitrogen in the interstellar medium.

**4 Developing ways to effectively use small telescopes**

4.1 Synchronous observations with distant telescopes

Ground-based networks of small and medium-sized optical telescopes are an important tool to investigate ultra-rapid variability of stellar brightness, to search for optical afterglow of gamma ray bursts, etc. because observations with a number of remote telescopes operating synchronously give information that cannot be obtained in any other way (Nogami et al. 2000; Zhilyaev et al. 2003).

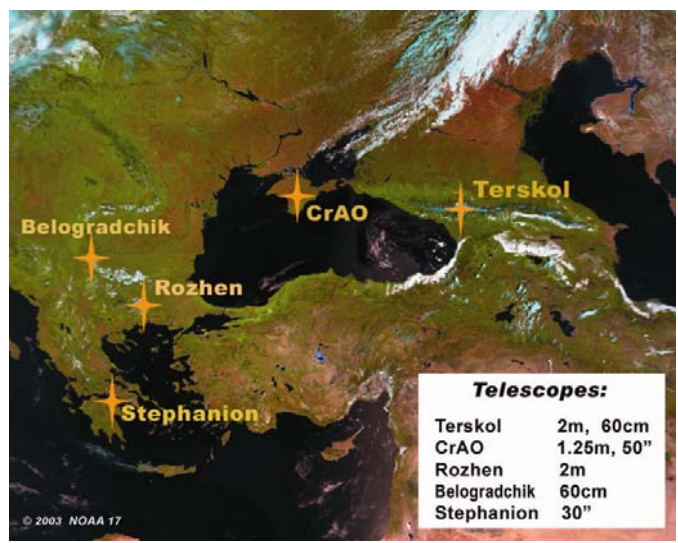

**Fig. 11** Observational sites and instruments of the Synchronous Network of distant Telescopes

The observations with the Synchronous Network of distant Telescopes (SNT), which involves telescopes in Ukraine, Russia, Bulgaria, and Greece, were started in the late 1990s (Fig. 11). The telescopes are equipped with GPS receivers to control local photometer timing systems relative to UTC. The high-speed, multi-site monitoring of variables, as well as digital filtering techniques used for data processing, provided new results:

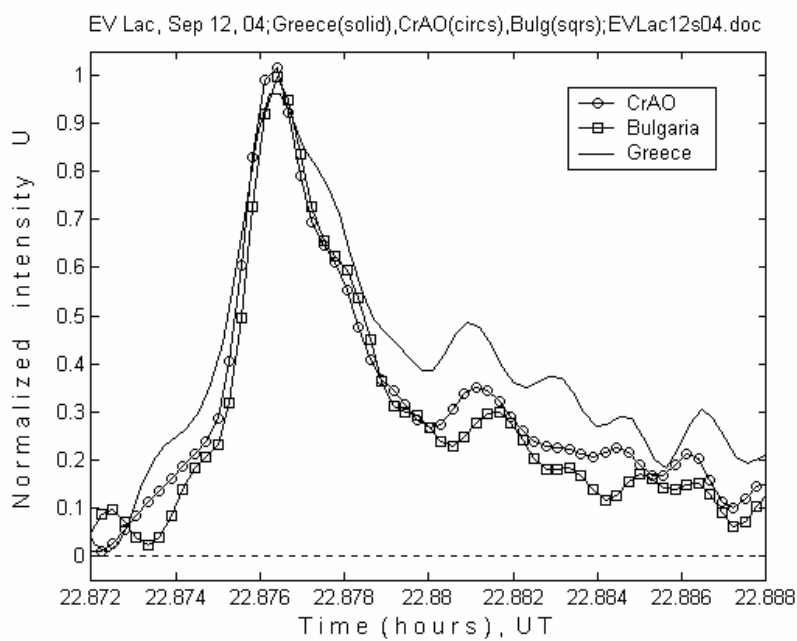

**Fig. 12** Multi-site photometry of a flare on EV Lac on September 12, 2004, 22:53 UT (max), as simultaneously seen by the three instruments: Ukraine (circles), Greece (solid), and Bulgaria (squares) (Zhilyaev et al. 2007)

(i) high-frequency oscillations in stellar flares were confirmed and (ii) fast color variations of the flare's radiation were discovered (Zhilyaev et al. 2007). Figure 12 demonstrates a flare on the star EV Lacertae, which was simultaneously observed with the three telescopes on September 12, 2004. Just the coincidence technique allows us to conclude that oscillations in stellar flares are connected with fast magnetoacoustic oscillations in coronal loops. This opens the way to study stellar coronae, i.e. their structure, density, and temperature (coronal seismology).

Furthermore, short burst events in some galaxies were revealed (Fig. 13). This observational result supports the hypothesis that intermediate-mass black holes exist in the centers of galaxies and dense globular clusters (Zhilyaev et al. 2006).

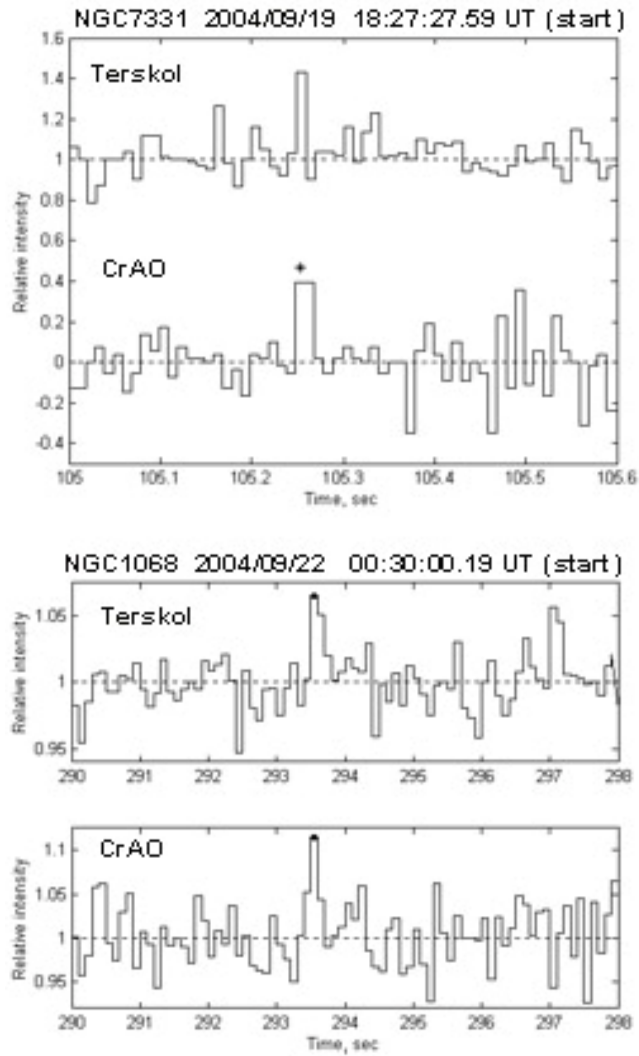

**Fig. 13** Detection of short-lived flare events in centers of the galaxies. *Top*: The light curves of NGC7331 taken on Sept 19, 2004, 18:27:27.59 UT (start time) synchronously at intervals of 10 ms in the B band using the Terskol 2-m telescope and the Crimean 50-inch telescopes. Both curves are in relative units, the lower one is shifted for convenience. *Bottom*: The same for the Seyfert galaxy NGC1068 (observations on Sept 22, 2004, 00:30:00.19 UT (start time)). The light curves are taken synchronously at intervals of 10 ms and rebinned to 0.5 s

4.2 A new observational and educational tool

The decade of successful research with SNT has yielded various analytical and numerical techniques to provide synchronous observations with distant telescopes. In 2006 the UNIT project (Ukrainian Network of Internet Telescopes) was initiated. It is aimed at the use of new technologies and systems to better demonstrate opportunities of modern astronomy and to create an interface between society and basic science.

The philosophy of UNIT is to develop an instrument to perform observations over the Internet from a PC at any location providing real-time access to data. The operations concept of UNIT would foster improvements in science and education (Godunova et al. 2008).

The system of UNIT, which is being developed, consists of three automatic telescopes. They are installed at two sites in Ukraine (a Celestron 14" near Kiev City and an 80-cm telescope near L'viv) and at the Terskol Observatory in the Northern Caucasus (a Celestron 11"). The number of telescopes in operation should increase in the near future due to other interested parties from Ukraine and abroad becoming involved.

The UNIT instruments are sensitive down to magnitude V ~ 18. For instance, they require about one minute to obtain the first images of a transient object after the alarm or GCN notice (slew speed up to 3° per second). The study of variables at magnitude U ~ 12 on a timescale of 1s could also be accomplished with UNIT. The telescopes are equipped with fast CCD cameras to study astrophysical events on timescales of up to tens of Hz. By using of GPS receivers, all exposures at the remote telescopes can be synchronized with an absolute accuracy of better than 1 ms. To observe transients, which are typically at magnitudes $10^m$ -$14^m$, we can use the coincidence technique for synchronous observations within UNIT and in that way obtain a time resolution of about 0.1 s.

**Conclusions**

Since the 1990s, the Terskol staff has put emphasis on the development and maintenance of astronomical facilities. We have come a long way, installing new instruments and developing new techniques and now we provide observational data of sufficient quality in many fields of astronomy. At present, the Terskol Observatory operates several optical telescopes with diameters up to 2 m. Successful scientific operation of the well-equipped telescopes and significant advances made over the past years have proven their usefulness for long-term observational programmes.

**Acknowledgements** First and above all, we wish to express our gratitude to many individuals who are or have been involved in multiple aspects of scientific and instrumental operation at the Terskol Observatory. We would like to express our sincere thanks to Prof. Klaus Jockers (Germany) for the excellent support he gave to the Terskol Observatory during the years 1996-2002. Prof. Jacek Krelowski (Poland) and Dr. Faig Musaev (Russia) are specially thanked for their hard work on the development of high-resolution spectrometric instrumentation at Terskol Peak. The collaboration of colleagues from the Institute of Astronomy (Russia) made the NEOs monitoring possible and is greatly appreciated.